\documentclass[aps,prl,twocolumn,superscriptaddress,floatfix]{revtex4-2}

\usepackage{graphicx}
\usepackage{dcolumn}
\usepackage{physics}
\usepackage{graphicx}
\usepackage{bm}
\usepackage{amssymb}
\usepackage{amsmath}
\usepackage{microtype}
\usepackage{xfrac}
\usepackage{array}
\usepackage[dvipsnames]{xcolor}
\usepackage[colorlinks,plainpages=false,linkcolor=blue,urlcolor=blue,citecolor=blue,pdfpagemode=UseNone,pdfstartview=FitBH]{hyperref}
\usepackage{nicematrix}

\newcommand{\angstrom}{\text{\normalfont\AA}} 

\begin{document}

\title{Zero-Field Long Range Order at $T \sim 40$~mK\\ in the Proximate Quantum Spin Ice \texorpdfstring{Ce$_2$Sn$_2$O$_7$}~\\ and Phase Diagram for Magnetic Fields Along $[1,1,0]$~}

\author{E.~M.~Smith}
\affiliation{Department of Physics and Astronomy, McMaster University, Hamilton, Ontario L8S 4M1, Canada}
\affiliation{Brockhouse Institute for Materials Research, McMaster University, Hamilton, Ontario L8S 4M1, Canada}

\author{M.~S.~Powell}
\affiliation{Department of Chemistry, Clemson University, Clemson, South Carolina 29634, United States}

\author{R.~Sch\"{a}fer}
\affiliation{Helmholtz-Zentrum Berlin f\"ur Materialien und Energie, Hahn-Meitner-Platz 1, 14109 Berlin, Germany}
\affiliation{Dahlem Center for Complex Quantum Systems and Fachbereich Physik, Freie Universit\"at Berlin, Arnimallee 14, 14195 Berlin, Germany}
\affiliation{Department of Physics, Harvard University, Cambridge, MA 02138, USA}

\author{A.~P.~Dioguardi}
\affiliation{Los Alamos National Laboratory, Los Alamos, New Mexico 87545, USA}

\author{J.~W.~Kolis}
\affiliation{Department of Chemistry, Clemson University, Clemson, South Carolina 29634, United States}

\author{R.~Movshovich}
\affiliation{Los Alamos National Laboratory, Los Alamos, New Mexico 87545, USA}

\author{B.~D.~Gaulin}
\affiliation{Department of Physics and Astronomy, McMaster University, Hamilton, Ontario L8S 4M1, Canada}
\affiliation{Brockhouse Institute for Materials Research, McMaster University, Hamilton, Ontario L8S 4M1, Canada}
\affiliation{Canadian Institute for Advanced Research, 661 University Avenue, Toronto, Ontario M5G 1M1, Canada.}

\date{\today}

\begin{abstract} 

The Ce$^{3+}$ pseudospin-1/2 degrees of freedom in the pyrochlore magnets Ce$_2$X$_2$O$_7$, with $X$ = Zr, Hf, or Sn, possess dipole-octupole character. The XYZ nearest-neighbor model Hamiltonians which have successfully described their properties make them attractive candidates for quantum spin ice ground states. We report new heat capacity measurements to very low temperatures on a high-quality single crystal of Ce$_2$Sn$_2$O$_7$ grown by hydrothermal techniques. Our zero-field measurements uncover a clear first order transition to long-ranged order at $T \sim 0.04$~K, well below the downturn in the broader Schottky-like anomaly that is a common feature in the zero-field heat capacity for cerium pyrochlores. This observation settles the debate as to whether Ce$_2$Sn$_2$O$_7$ possesses a QSI ground state in zero field - it does not. However, we also find compelling evidence suggesting that spin ice physics is nearby, and remains relevant to Ce$_2$Sn$_2$O$_7$. Application of a magnetic field along the $[1,1,0]$ direction leads to an evolution of this peak to a weak anomaly at higher temperature more characteristic of a continuous, mean field transition. At higher fields along $[1,1,0]$, the Schottky-like anomaly bifurcates similar to expectations for independent polarized $\alpha$ and orthogonal $\beta$ chains in classical spin ice. These new experimental results demonstrate richness to the phase diagram for Ce-based pyrochlores. 
\end{abstract}
\maketitle

Quantum spin ices (QSIs) are a specific type of quantum spin liquid in which the spin disorder mimics the proton disorder in water ice and the spin interaction Hamiltonian maps on to an emergent quantum electrodynamics where the exotic elementary excitations correspond to magnetic and electric monopoles as well as emergent photons~\cite{Hermele2004, Banerjee2008, Lee2012, Benton2012, Savary2012b, GingrasReview2014}. Cubic pyrochlores, comprised of networks of corner-sharing tetrahedra, possess a similar tetrahedral coordination as oxygen ions in water ice and are a natural venue for spin ice physics. The Ce-based insulating pyrochlores, Ce$_2$Zr$_2$O$_7$, Ce$_2$Hf$_2$O$_7$, and Ce$_2$Sn$_2$O$_7$ have drawn much recent attention as candidates for QSI ground states. In this environment, crystal electric field (CEF) effects break the $J=5/2$ Hund's rule ground state associated with Ce$^{3+}$ into three well separated doublets~\cite{Sibille2015, Gaudet2019, Gao2019, Sibille2020, Poree2022}, and yield a pseudospin-$1/2$ degree of freedom for each Ce$^{3+}$ ion~\cite{Curnoe2007, Onada2011, Huang2014, RauReview2019}. The wavefunctions associated with the CEF ground state doublet correspond to a $z$-component of pseudospin with a dipole moment, which transforms as a dipole under time-reversal symmetry and the point group symmetry at the Ce-site. The $x$ and $y$ components of pseudospin carry octupole moments but transform differently from each other under time-reversal symmetry and the point group symmetry at the Ce-site, with the $x$ component transforming as a dipole and the $y$ component transforming as an octupole~\cite{Huang2014, Li2017, RauReview2019, Smith2025a}. Such materials are known as dipole-octupole magnets.

The symmetry of the CEF ground state also dictates the general form of the pseudospin interaction Hamiltonian~\cite{RauReview2019}. The symmetry-allowed pseudospin-1/2 interaction Hamiltonian at the nearest-neighbor level for pyrochlores with a dipole-octupole CEF ground state doublet is given by~\cite{Huang2014, Li2017, RauReview2019, Smith2025a}:

\begin{equation}\label{eq:1}
\begin{split}
    \mathcal{H}_\mathrm{DO} & = \sum_{\langle ij \rangle}[J_{x}{S_i}^{x}{S_j}^{x} + J_{y}{S_i}^{y}{S_j}^{y} + J_{z}{S_i}^{z}{S_j}^{z} \\ 
    & + J_{xz}({S_i}^{x}{S_j}^{z} + {S_i}^{z}{S_j}^{x})] - g_z \mu_\mathrm{B} \sum_{i} \mathbf{h} \cdot\hat{{\bf z}}_i \; {S_i}^{z}  \;,
\end{split}
\end{equation}

\noindent where ${S_{i}}^{x}$, ${S_{i}}^{y}$, and ${S_{i}}^{z}$ are the pseudospin components of the rare-earth atom $i$ in the local $\{x$, $y$, $z\}$ coordinate frame. This coordinate frame is defined locally for each ion $i$ with the $\mathbf{z}_i$ anisotropy axis along the threefold rotation axis through rare-earth site $i$ and with $\mathbf{y}_i$ along one of the symmetrically equivalent twofold rotation axes through rare-earth site $i$, where $\mathbf{x}_i = \mathbf{y}_i \cross \mathbf{z}_i$~\cite{Ross2011,RauReview2019}. These local axes are defined in terms of the global coordinate frame in the Supplemental Material~\cite{SM}. The last sum in Eq.~\ref{eq:1} represents the Zeeman interaction between the magnetic field $\mathbf{h}$ and the magnetic dipole moments associated with the $S^z$ components of pseudospin. The anisotropic $g$-factor has only a non-zero $g_z$ component determined by the CEF ground state doublet for Ce$^{3+}$, estimated to be $g_z \approx 2.2$ for Ce$_2$Sn$_2$O$_7$~\cite{Sibille2015, Yahne2024}.

This nearest-neighbor exchange Hamiltonian can then be simplified via rotation of each local $\{x, y, z\}$ coordinate frame by $\theta$ about the respective local $y$-axis, where $\theta$ is given by~\cite{Huang2014, Benton2016}:
\begin{equation} \label{eq:2}
    \theta = \frac{1}{2}\tan^{-1}\bigg(\frac{2J_{xz}}{J_{x}-J_{z}}\bigg)\;.
\end{equation}

These rotations yield new local coordinate frames which are commonly denoted as the local $\{\tilde{x},\tilde{y},\tilde{z}\}$ coordinate frames, and the new Hamiltonian in the $\{\tilde{x},\tilde{y},\tilde{z}\}$ coordinate frames is the ``XYZ'' Hamiltonian~\cite{Huang2014, Benton2016, Smith2025a}:

\begin{equation} \label{eq:3}
\begin{split}
    \mathcal{H}_\mathrm{XYZ} & = \sum_{\langle ij \rangle}[     J_{\tilde{x}}{S_i}^{\tilde{x}}{S_j}^{\tilde{x}} + J_{\tilde{y}}{S_i}^{\tilde{y}}{S_j}^{\tilde{y}} + J_{\tilde{z}}{S_i}^{\tilde{z}}{S_j}^{\tilde{z}}] \\ 
    & - g_z \mu_\mathrm{B} \sum_{i} \mathbf{h} \cdot\hat{{\bf z}}_i({S_i}^{\tilde{z}}\cos\theta + {S_i}^{\tilde{x}}\sin\theta).
\end{split}
\end{equation}

This nearest-neighbor model for the dipole-octupole pyrochlores permits both QSI and ordered ground states, where each can have either a dipolar or octupolar flavor~\cite{Huang2014, Benton2020, Patri2020, Huang2020}.

All of Ce$_2$Zr$_2$O$_7$, Ce$_2$Hf$_2$O$_7$, and Ce$_2$Sn$_2$O$_7$ have been studied experimentally, using both powder and single crystal samples~\cite{Gao2019, Gao2022, Changlani2022, Gaudet2019, Sibille2015, Sibille2020, Yahne2024, Poree2022, Poree2025a, Poree2025b, Smith2022, Beare2023, Smith2023, Smith2024, Smith2025a, Smith2025b, Gao2025, Bhardwaj2025, Kermarrec2025, Yuan2026}.  The single crystals of Ce$_2$Zr$_2$O$_7$ and Ce$_2$Hf$_2$O$_7$ studied to date were grown from the melt by floating zone image furnace techniques at temperatures exceeding 2000~$^{\circ}$C.  Powder samples of all three were grown by solid state synthesis using somewhat lower temperatures of $\sim$ 1000~$^{\circ}$C.  A recent breakthrough in single crystal synthesis has employed hydrothermal techniques and can be achieved using lower temperatures still: 700~$^{\circ}$C~\cite{Powell2019}. Following the hydrothermal synthesis procedure described for Ce$_2$Sn$_2$O$_7$ in Ref.~\cite{Powell2019}, we have grown small single crystals of Ce$_2$Sn$_2$O$_7$ which have enabled recent diffuse neutron scattering measurements in Ref.~\cite{Yuan2026} as well as the present thermodynamic measurements. The recent diffuse scattering measurements on single crystals of Ce$_2$Sn$_2$O$_7$ show sharper diffuse scattering features than those of the Ce$_2$Zr$_2$O$_7$ single crystal~\cite{Gao2019, Gaudet2019, Smith2022}. This observation was attributed to much reduced defect concentrations in the hydrothermally grown Ce$_2$Sn$_2$O$_7$, compared to floating-zone grown single crystals of Ce$_2$Zr$_2$O$_7$, due to the lower growth-temperature.

In this letter we report on heat capacity measurements performed on a 44.1~mg single crystal of Ce$_2$Sn$_2$O$_7$ following the quasi-adiabatic procedure described for Ce$_2$Hf$_2$O$_7$ in Ref.~\cite{Smith2025b}. Our measurements for non-zero magnetic field used a field-cooled protocol, cooling at the same field strength as used for the corresponding measurement in each case. A low base-temperature (0.022~K in the zero-field case) is possible due to the relatively large sample mass used for the measurements, allowing for a strong thermal linkage between the sample and the dilution refrigerator. Furthermore, the large relaxation time-constant of Ce$_2$Sn$_2$O$_7$ at low temperature allowed for careful equilibration protocols with relatively slow heat pulses and a long averaging-time for thermometer readings, leading to high precision and equilibrated measurements at very low temperatures. 


\begin{figure}[t]
\linespread{1}
\par
\includegraphics[width=3.4in]{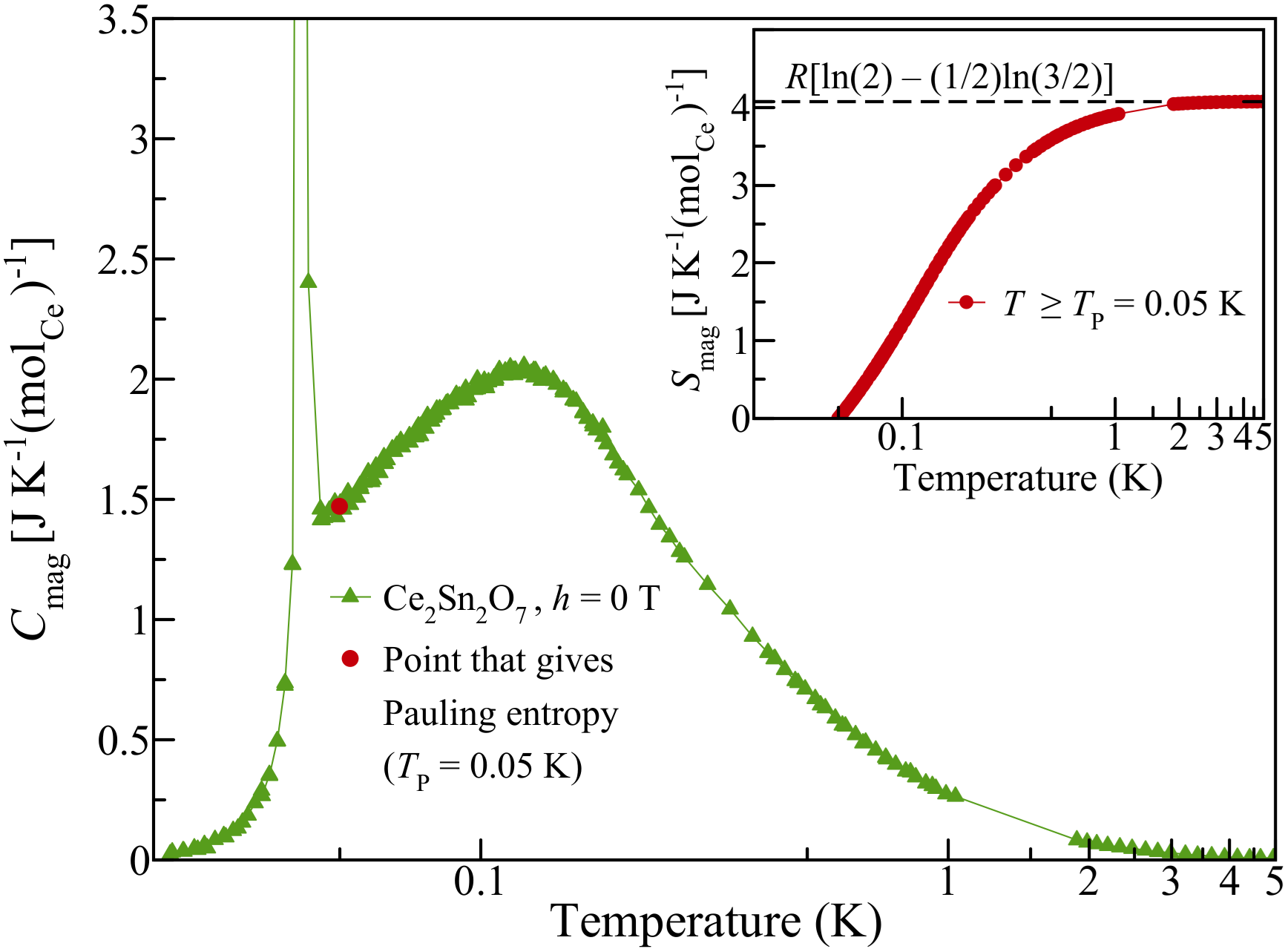}
\par
\caption{The zero-field heat capacity of hydrothermally-grown single crystal Ce$_2$Sn$_2$O$_7$. A large, discontinuous phase transition is obvious at $T_N \sim 0.04$~K; This peak is more than three orders of magnitude higher than shown [see Fig.~\ref{Figure3}(a)].  The red circle just above the anomaly signifies the point above which the measured entropy gives the Pauling entropy $R[\ln(2)-(1/2)\ln(3/2)]$, indicating a fully developed spin ice state above $T_N$. The inset shows the entropy recovered above the aforementioned ``Pauling point'', $T_{\mathrm{P}} = 0.05$~K, in the zero-field heat capacity of Ce$_2$Sn$_2$O$_7$ via $S_\mathrm{mag} = \int_{T_{P}}^T\frac{C_\mathrm{mag}}{T}\mathrm{d}T$.}
\label{Figure1}
\end{figure}

\begin{figure}[t]
\linespread{1}
\par
\includegraphics[width=3.4in]{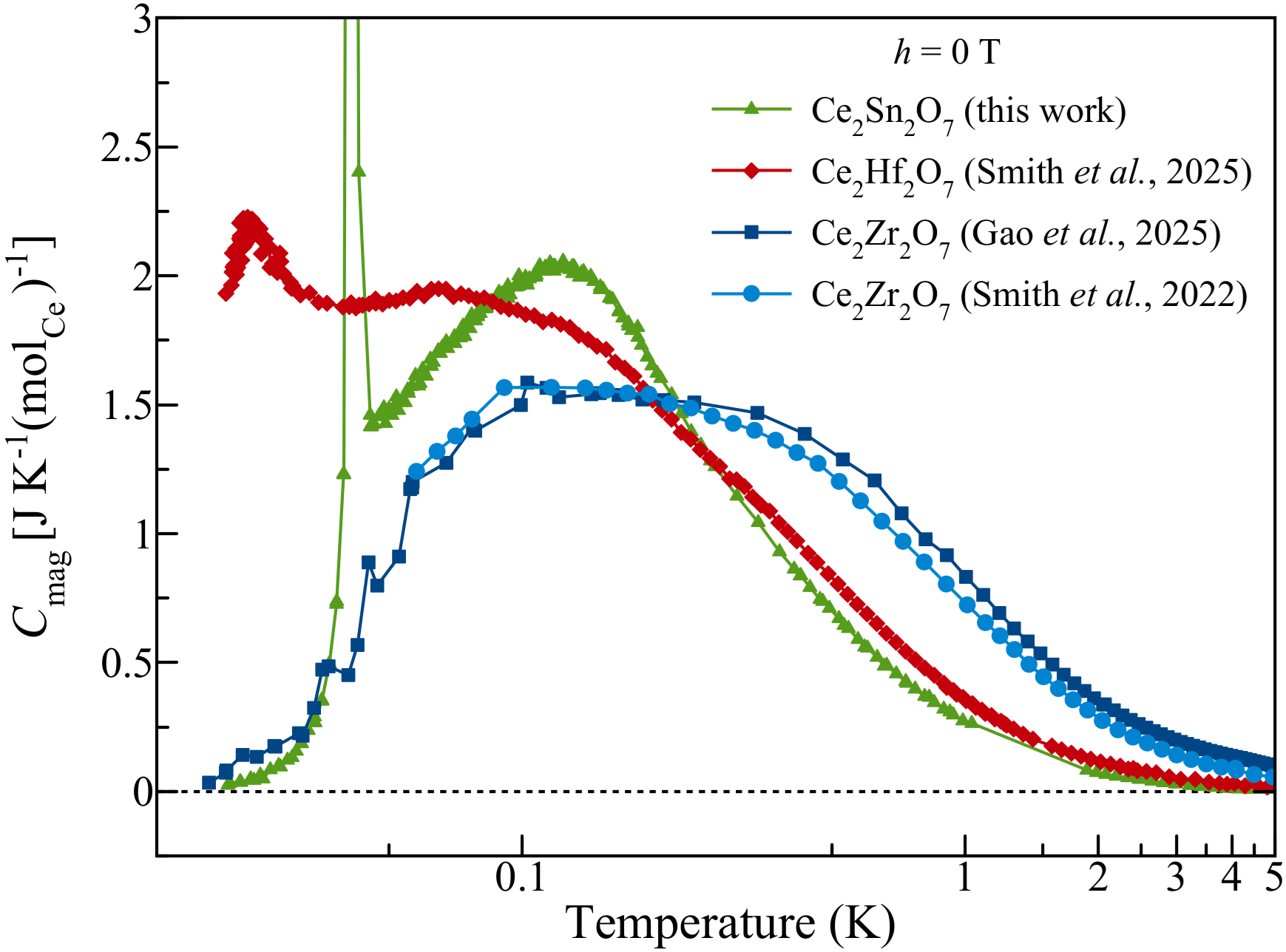}
\par
\caption{The zero-field heat capacity measured from the hydrothermally-grown Ce$_2$Sn$_2$O$_7$ single crystal in this work (green) compared with that measured from floating-zone-grown single crystals of Ce$_2$Hf$_2$O$_7$ in Ref.~\cite{Smith2025b} (red) and Ce$_2$Zr$_2$O$_7$ in Refs.~\cite{Smith2022,Gao2025} (light blue, dark blue).} 
\label{Figure2}
\end{figure}

Figure~\ref{Figure1} shows the zero field heat capacity of Ce$_2$Sn$_2$O$_7$ measured between $T = 0.022$~K and 5~K on a logarithmic temperature scale and a linear $C_{mag}$ scale. Measurements on La$_2$Zr$_2$O$_7$ and other non-magnetic pyrochlores have indicated that lattice contributions are negligible for $T < 5$~K ~\cite{Saha2008, Krizan2015, Xu2015, Smith2022, Smith2023}. This C$_{mag}$ dataset shows two well-separated anomalies: a huge, sharp anomaly near $T_N \sim 0.04$~K and a higher-temperature Schottky-like anomaly just above $T = 0.1$~K.  The Schottky-like anomaly is very similar to that observed in the classical spin ices Ho$_2$Ti$_2$O$_7$ and Dy$_2$Ti$_2$O$_7$~\cite{Bramwell2001, Higashinaka2003, Gronemann2023}, and to that observed in the quantum spin ice candidates Ce$_2$Zr$_2$O$_7$ and Ce$_2$Hf$_2$O$_7$ as can be seen in Fig.~\ref{Figure2}~\cite{Gao2019, Smith2022, Smith2023, Gao2025, Poree2022, Poree2025b}. Here the temperature dependence of the zero-field $C_{mag}$ data for single crystals of all three of Ce$_2$Zr$_2$O$_7$, Ce$_2$Sn$_2$O$_7$ and Ce$_2$Hf$_2$O$_7$ are overplotted, again on a logarithmic temperature scale and a linear $C_{mag}$ scale. While a Schottky-like feature is observed in $C_{mag}$ at moderate temperatures in all three Ce-based pyrochlores, the large anomaly for Ce$_2$Sn$_2$O$_7$, at $T_N \sim 0.04$~K, is qualitatively new and signifies a discontinuous phase transition to some magnetically ordered phase. 

As the two anomalies are well-separated in temperature and the low-temperature anomaly is so sharp, one can ask how much entropy is released above the discontinuous phase transition at $T_N \sim 0.04$~K. The inset to Fig.~\ref{Figure1} shows the entropy recovered starting from the ``Pauling point'' at $T_{\mathrm{P}} = 0.05$~K and going to high temperatures (5~K), which by design accounts for the recovered entropy in a classical spin ice, $R[\ln(2) - (1/2)\ln(3/2)]$~\cite{Pauling1935, Anderson1956, Harris1997, Ramirez1999, Bramwell2001}. Together with the prior observation of a spin ice phase at temperatures above $T_N \sim 0.04$~K via neutron diffraction measurements on hydrothermally-grown Ce$_2$Sn$_2$O$_7$~\cite{Yahne2024, Yuan2026}, this makes the case that Ce$_2$Sn$_2$O$_7$ is in a fully formed classical spin ice state at the ``Pauling point" $T_{\mathrm{P}} = 0.05$~K, just above~$T_N$. 

The magnetic ordering transition at $T_N \sim 0.04$~K is consistent with with expectations based on the estimated exchange parameters and the quantum Monte Carlo simulations for Ce$_2$Sn$_2$O$_7$ in Ref.~\cite{Yahne2024} using the nearest neighbor XYZ Hamiltonian, which together predict a transition around $T_N \sim 0.04$~K to an all-in all-out zero-field ground state with ordered ${S}^{\tilde{z}}$ components of pseudospin. This contrasts with Refs.~\cite{Sibille2020, Poree2025a}, where measurements on powder Ce$_2$Sn$_2$O$_7$ are interpreted as Ce$_2$Sn$_2$O$_7$ having an octupolar quantum spin ice ground state.

The presence of a classical spin ice state at the ``Pauling point" just above $T_N$ is also consistent with expectations based on the estimates of the XYZ Hamiltonian parameters for Ce$_2$Sn$_2$O$_7$ in Ref.~\cite{Yahne2024}, $(J_{\tilde{x}}, J_{\tilde{y}}, J_{\tilde{z}}) = (0.045, -0.001, -0.012)$~meV and $\theta = 0.19\pi$, which place Ce$_2$Sn$_2$O$_7$ in a region of parameter space where an ${S}^{\tilde{x}}$-flavored spin ice phase is classically stable in zero field but is ultimately unstable to the all-in all-out zero-field ground state when quantum effects are considered. However, despite this consistency with  expectations based on the nearest-neighbor XYZ Hamiltonian~\cite{Yahne2024}, the recent neutron scattering work on hydrothermally grown Ce$_2$Sn$_2$O$_7$ in Ref.~\cite{Yuan2026} finds no signs of short-ranged all-in all-out order at $T = 0.05$~K (slightly above $T_N$) and uncovers features of the low-temperature diffuse scattering that are suggestive of significant long-ranged dipole-dipole interactions. Together these findings of Ref.~\cite{Yuan2026} suggest that the nearest-neighbor XYZ Hamiltonian alone provides an incomplete description of the low temperature ($T \lesssim 0.05$~K) physics in Ce$_2$Sn$_2$O$_7$.

\begin{figure}[t]
\linespread{1}
\par
\includegraphics[width=3.4in]{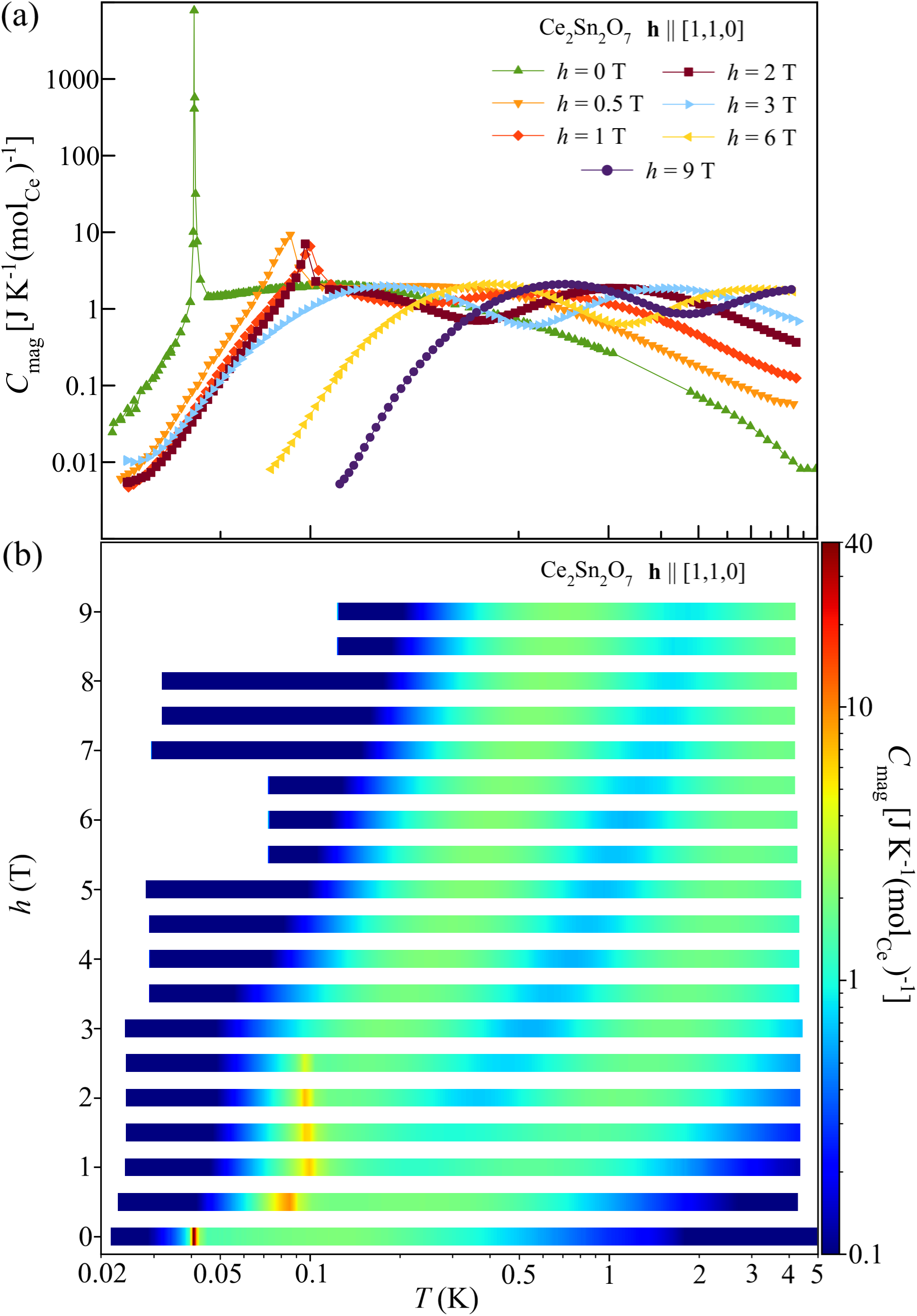}
\par
\caption{(a) The temperature-dependence of the heat capacity of Ce$_2$Sn$_2$O$_7$ at various field strengths (as labeled) for a magnetic field $\mathbf{h}$ along the $[1,1,0]$ direction, shown with logarithmic temperature and heat capacity axes. (b) The magnetic field versus temperature phase diagram for Ce$_2$Sn$_2$O$_7$ in a magnetic field nominally along the $[1,1,0]$ direction, as directly determined from the $C_\mathrm{mag}$ measurements in (a). The data at each field strength have been given an artificial width in field-strength for visibility. The dataset shown for 1.5~T (2.5~T, 3.5~T, 4.5~T, 5.5~T, 6.5~T, 7.5~T, 8.5~T) results from averaging the measured data at 1 and 2~T (2 and 3~T, 3 and 4~T, 4 and 5~T, 5 and 6~T, 6 and 7~T, 7 and 8~T, 8~and~9~T).} 
\label{Figure3}
\end{figure}

\begin{figure}[!h]
\linespread{1}
\par
\includegraphics[width=3.4in]{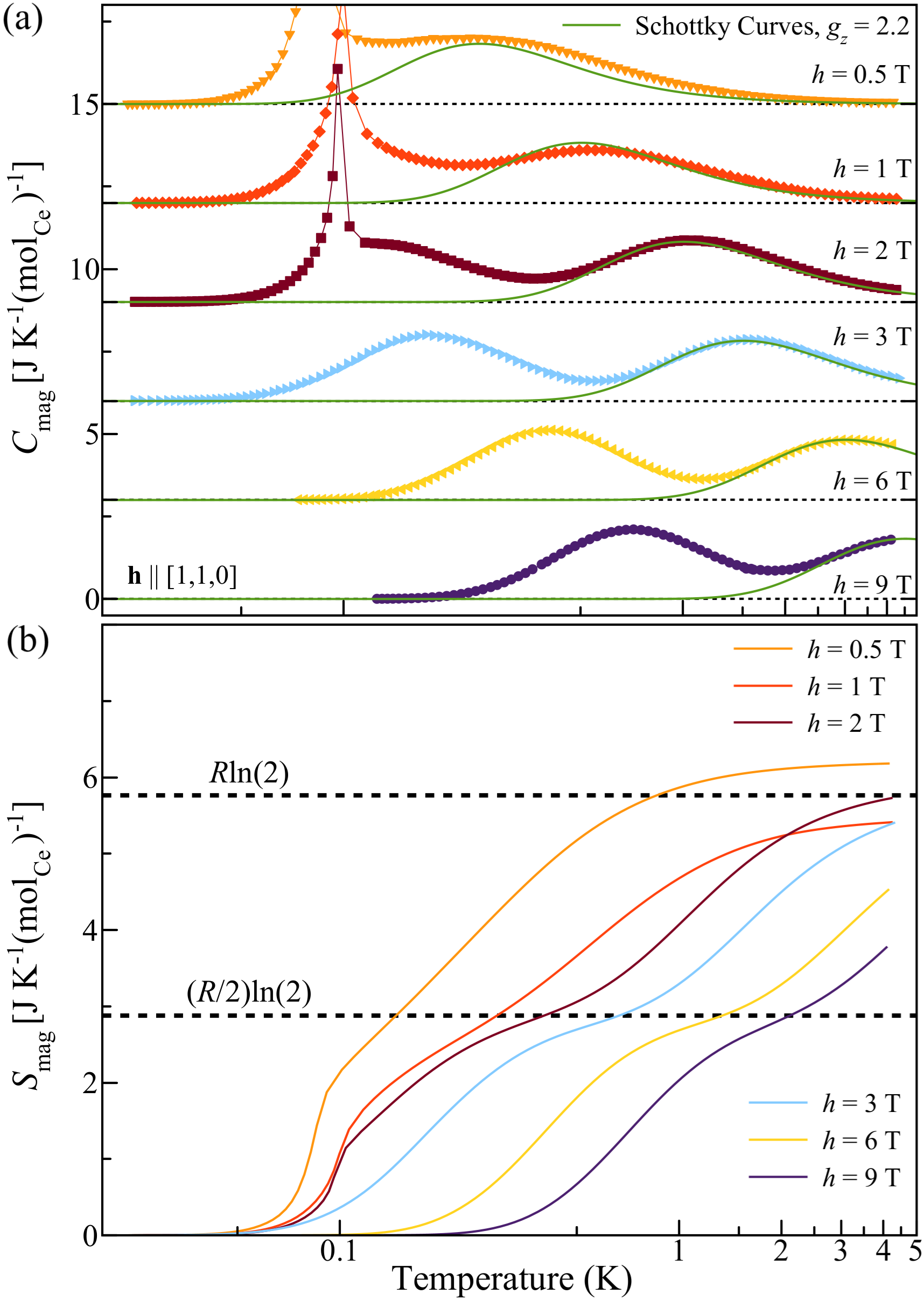}
\par
\caption{(a) The temperature-dependence of the heat capacity of Ce$_2$Sn$_2$O$_7$ at various field strengths (as labeled) for a magnetic field along the $[1,1,0]$ direction, shown with a logarithmic temperature axis and linear heat capacity axis with the data offset by 3~J~K$^{-1}$mol$_{\mathrm{Ce}}^{-1}$ between each pair of successive field strengths. We compare these measurements to the curves predicted according to the two-level Schottky form $C_{\alpha} = \frac{R}{2}(\frac{\Delta}{2k_{\mathrm{B}}T})^2 \sech^2(\frac{\Delta}{2k_{\mathrm{B}}T})$ where $\Delta = g_z \mu_{\mathrm{B}} h_z$ is the gap associated with the Zeeman splitting for a single ion within the $\alpha$ chains~\cite{Hiroi2003}, and $h_z = h\cos(35.26^\circ)$ for the $\alpha$-chain ions in a field along $[1,1,0]$. Here we use the value of $g_z = 2.2$ estimated for Ce$_2$Sn$_2$O$_7$ in Refs.~\cite{Sibille2015, Yahne2024}. (b)~The entropy recovered above base temperature and up to $T = 5$~K for Ce$_2$Sn$_2$O$_7$ via $S_\mathrm{mag} = \int_{0}^T\frac{C_\mathrm{mag}}{T}\mathrm{d}T$ using the $C_\mathrm{mag}$ measurements in (a).}
\label{Figure4}
\end{figure}

In Figs.~\ref{Figure3} and \ref{Figure4} we now shift to $C_\mathrm{mag}$ measurements with a magnetic field applied along the $[1,1,0]$ direction in our single crystal of Ce$_2$Sn$_2$O$_7$. Figure~\ref{Figure3}(a) shows these $C_\mathrm{mag}$ datasets plotted with logarithmic $C_\mathrm{mag}$ and $T$ axes for $\mathbf{h} \parallel [1,1,0]$ between $h = 0$ and 9~T.  With $C_\mathrm{mag}$ on a log scale, we can now see the extent to which the zero-field anomaly at $ \sim 0.04$~K is both sharp in temperature and very large in magnitude. This low-temperature zero-field anomaly weakens and moves up in temperature to about 0.09~K, for $h$ between $\sim 0.5$~T and $\sim2$~T, and is thereafter not obvious, as it merges into the broad Schottky-like features present in finite $\mathbf{h} \parallel [1,1,0]$.   

The same $C_\mathrm{mag}$ data in Fig.~\ref{Figure3}(a) are shown as a color plot in Fig.~\ref{Figure3}(b), where it maps out an approximate $h$~vs.~$T$ phase diagram. One clearly sees that the $[1,1,0]$ magnetic field also bifurcates the zero-field Schottky-like anomaly, producing two Schottky-like anomalies that smoothly connect to the zero-field $C_\mathrm{mag}$. The magnetic Ce$^{3+}$-sublattice in Ce$_2$Sn$_2$O$_7$ can be thought of as being comprised of orthogonal $\alpha$ and $\beta$ chains that are decoupled from one another when the $\alpha$ chains along $[1,1,0]$ are polarized by a magnetic field in that direction~\cite{Yoshida2004, Placke2020}. Indeed, our results can be interpreted as the polarization of the dipole moments along the local $z$ directions in the $\alpha$ chains, while the dipole moments on the $\beta$ chains are nominally orthogonal to the field direction such that the $\beta$ chains are less affected by the field. Such a scenario is very well established in classical spin ices like Dy$_2$Ti$_2$O$_7$~\cite{Fennell2002, Hiroi2003, Fennell2005, Ruff2005} and Ho$_2$Ti$_2$O$_7$~\cite{Harris1997, Fennell2005, Clancy2009}, and has recently been investigated in Ce$_2$Zr$_2$O$_7$ and Ce$_2$Hf$_2$O$_7$~\cite{Gao2022, Smith2023, Beare2023, Zhou2024, Bhardwaj2025, Kermarrec2025}.


The evolution of the Schottky-like features in the $C_{mag}$ of Ce$_2$Sn$_2$O$_7$ in a $[1,1,0]$ magnetic field are more-easily followed using a linear $C_{mag}$, as shown in Fig.~\ref{Figure4}(a).  The bifurcation into two broad Schottky-like forms is clear, with the upper-temperature feature associated with the $\alpha$ chains, as it shows stronger $h$-dependence due to the direct non-colinear polarization of the $\alpha$ chains by the $[1,1,0]$ field. It is interesting to note that the sharp low-temperature anomaly, at $T_N \sim$ 0.04~K in zero field, is only evident so long as the field is not sufficiently strong to separate the two Schottky-like anomalies.  This occurs around $h=2$~T, and indeed no sharp $C_{mag}$ anomaly is observed for $h=3$~T or higher. This suggests that the low-field ground state cannot survive the decomposition of the pyrochlore lattice into $\alpha$ and $\beta$ chains at higher $[1,1,0]$ fields.
 
The upper-temperature Schottky anomaly of Ce$_2$Sn$_2$O$_7$ in a magnetic field along $[1,1,0]$ can be quantitatively modeled assuming only that it arises from the polarization of the $\alpha$ chains, such that the component of the $[1,1,0]$ field along the local $z$-direction for the Ce$^{3+}$ pseudospin is given by $h_z = h\cos(35.26^\circ)$. The high temperature Schottky term in the C$_{mag}$ data in Fig.~\ref{Figure4}(a) is then modeled by $C_{\alpha} = \frac{R}{2}(\frac{\Delta}{2k_{\mathrm{B}}T})^2 \sech^2(\frac{\Delta}{2k_{\mathrm{B}}T})$~\cite{Hiroi2003}, where $\Delta = g_z\mu_{\mathrm{B}}h_z$ and we use $g_z = 2.2$ as estimated for Ce$_2$Sn$_2$O$_7$ in Refs.~\cite{Sibille2015, Yahne2024}. The resulting Schottky curve $C_{\alpha}$ is shown as the solid line in Fig.~\ref{Figure4}(a) and clearly the description of the high temperature data is excellent, especially where the two broad anomalies are well separated ($h = 2$~T and beyond). Accordingly, we associate this higher-temperature hump with the polarization of the $\alpha$ chains and the lower-temperature hump with the behavior of the $\beta$ chains. 

Figure~\ref{Figure4}(b) shows the recovery of entropy with temperature as a function of [1,1,0] magnetic field, based on the $C_{mag}$ data in Fig.~\ref{Figure4}(a) above base temperature and up to $T = 5$~K using $S_\mathrm{mag} = \int_{0}^T\frac{C_\mathrm{mag}}{T}\mathrm{d}T$.  What is interesting, but perhaps expected, is the clear observation of plateau-like features at $(R/2)\ln(2)$ for all non-zero values of $\mathbf{h} \parallel [1,1,0]$, half of the full $R\ln(2)$ expected for the pseudospin-1/2 degrees of freedom. This is naturally associated with the fact the $\alpha$ and $\beta$ chains possess different energy scales, such that upon decreasing temperature, a temperature is reached at which the $\alpha$ chains are near their full non-collinear polarization with $\sim 0$ entropy while the $\beta$ chains, with the other half of the pseudospins in the system, remain paramagnetic with $\sim (R/2)\ln(2)$ entropy until lower temperature.

We note that the lower-temperature hump in the heat capacity is not fit well by a two-level Schottky form for any gap energy $\Delta$, even at high fields, suggesting that more-complex physics is at play for the $\beta$ chains. The XYZ Hamiltonian parameters estimated for Ce$_2$Sn$_2$O$_7$ in Ref.~\cite{Yahne2024}, with $J_{\tilde{x}} \gg |J_{\tilde{y}}|$,$|J_{\tilde{z}}| > 0$, suggest that antiferromagnetic intrachain correlations between ${S}^{\tilde{x}}$ components of pseudospin would be the dominant correlation in the $\beta$ chains of Ce$_2$Sn$_2$O$_7$ in moderate and high $[1,1,0]$-fields. These correlations are ferromagnetic with respect to a \textit{global} coordinate system. In light of the diffuse neutron scattering in Ref.~\cite{Yuan2026} highlighting the significance of long-ranged dipole-dipole interactions in Ce$_2$Sn$_2$O$_7$, it is likely that dipole-dipole interactions play a large role in any interchain coupling of the $\beta$ chains. In addition to this, it is expected that even a small misalignment of the magnetic field away from the $[1,1,0]$ direction (as is unavoidable in experiment) leads to a significant coupling between the magnetic field and the $\beta$ chains, as has been found for Ce$_2$Hf$_2$O$_7$ in Ref.~\cite{Bhardwaj2025}.  

To conclude, we report $C_{mag}$ measurements on a new hydrothermally grown single crystal of Ce$_2$Sn$_2$O$_7$ down to $T \sim 0.02$~K. These measurements have discovered a first-order transition to a long range ordered state at $T \sim 0.04$~K, settling a longstanding debate as to whether the magnetic ground state is ordered or disordered in high-quality Ce$_2$Sn$_2$O$_7$. While this ordered ground state has yet to be identified experimentally, the presence of a long-ranged ordered state below $T \sim 0.04$~K is consistent with expectations based on prior estimates for the exchange parameters in the nearest-neighbor XYZ Hamiltonian for Ce$_2$Sn$_2$O$_7$~\cite{Yahne2024}. Under application of a $[1,1,0]$ magnetic field, Ce$^{3+}$ pseudospins on the $\alpha$ and $\beta$ chains that make up the pyrochlore lattice evolve differently such that each contribute their own Schottky-like anomaly for $h \gtrsim 2$~T. The phase transition to long-range order does not survive the bifurcation of the pyrochlore lattice into $\alpha$ and $\beta$ chains induced by $\mathbf{h} \parallel [1,1,0]$, but comes to a novel end point just beyond $h = 2$~T. We hope these results will motivate experimental and theoretical work focused on a greater understanding of the rich physics in the Ce-based pyrochlores at low temperatures both in zero and non-zero magnetic fields.

\begin{acknowledgments}
This work was supported by the Natural Sciences and Engineering Research Council of Canada (NSERC). The synthesis and crystal growth of the material and partial analysis and data collection were supported by DoE grant DE-SC0020071. Work at Los Alamos National Laboratory was supported by the U.S. Department of Energy, Office of Science, National Quantum Information Science Research Centers, Quantum Science Center. R.~S. acknowledges support from the DFG under Project No. 575641691.
\end{acknowledgments}

%

\clearpage

\renewcommand{\thefigure}{S\arabic{figure}}
\setcounter{figure}{0}

\section{SUPPLEMENTAL MATERIAL:}

\section{Details of Heat Capacity Analysis}

\begin{figure}[t]
\linespread{1}
\par
\includegraphics[width=3.4in]{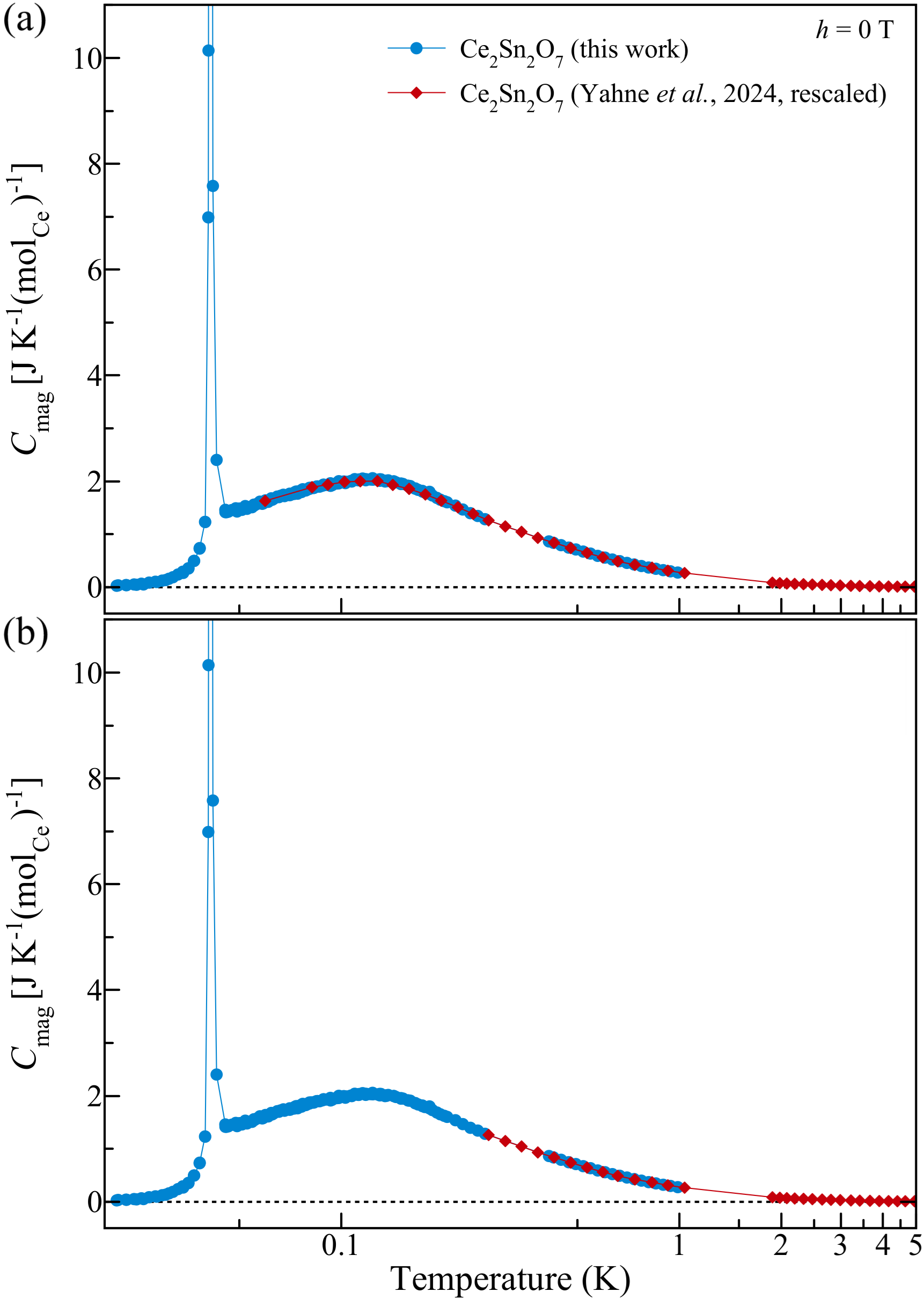}
\par
\caption{(a) A comparison of the zero-field heat capacity measured from single crystal Ce$_2$Sn$_2$O$_7$ in this work (blue) with that measured in Ref.~\cite{Yahne2024} (red). The data from Ref.~\cite{Yahne2024} has been re-scaled by 0.87 for best agreement with the new data, which has a negligible mass uncertainty compared to the $\sim$13\% mass uncertainty from Ref.~\cite{Yahne2024}. (b) The combined zero-field heat capacity dataset for Ce$_2$Sn$_2$O$_7$ formed by combining the measurements of the present work with the $T > 0.25$~K portion of the re-scaled data from Ref.~\cite{Yahne2024}. This combined dataset is the zero-field dataset used for Ce$_2$Sn$_2$O$_7$ elsewhere throughout this work.} 
\label{FigureS1}
\end{figure}

The heat capacity measured from single crystal Ce$_2$Sn$_2$O$_7$ in the present work is shown in Fig.~\ref{FigureS1}(a) where we compare the results with the zero-field heat capacity measured from (also hydrothermally-grown) single crystal Ce$_2$Sn$_2$O$_7$ in Ref.~\cite{Yahne2024}. The measurements in Ref.~\cite{Yahne2024} have systematic uncertainty of $\sim$ 13\% associated with uncertainty in the mass of the small sample used in that work, whereas the heat capacity measurements of the present work have a negligible mass uncertainty ($<0.1\%$). We find best agreement between our zero-field measurements and those in Ref.~\cite{Yahne2024} when scaling the data from Ref.~\cite{Yahne2024} by 0.87, as shown in Fig.~\ref{FigureS1}(a). We use this scaling of the zero-field data in Ref.~\cite{Yahne2024} to connect our measurements to the re-scaled data from Ref.~\cite{Yahne2024} and the combined dataset that we use throughout this work is shown in Fig.~\ref{FigureS1}(b). Specifically, we use the combined dataset that results from combining the $T > 0.25$~K portion of the data from Ref.~\cite{Yahne2024} with the full zero-field dataset of our new measurements.

\begin{figure*}[t]
\linespread{1}
\par
\includegraphics[width=7in]{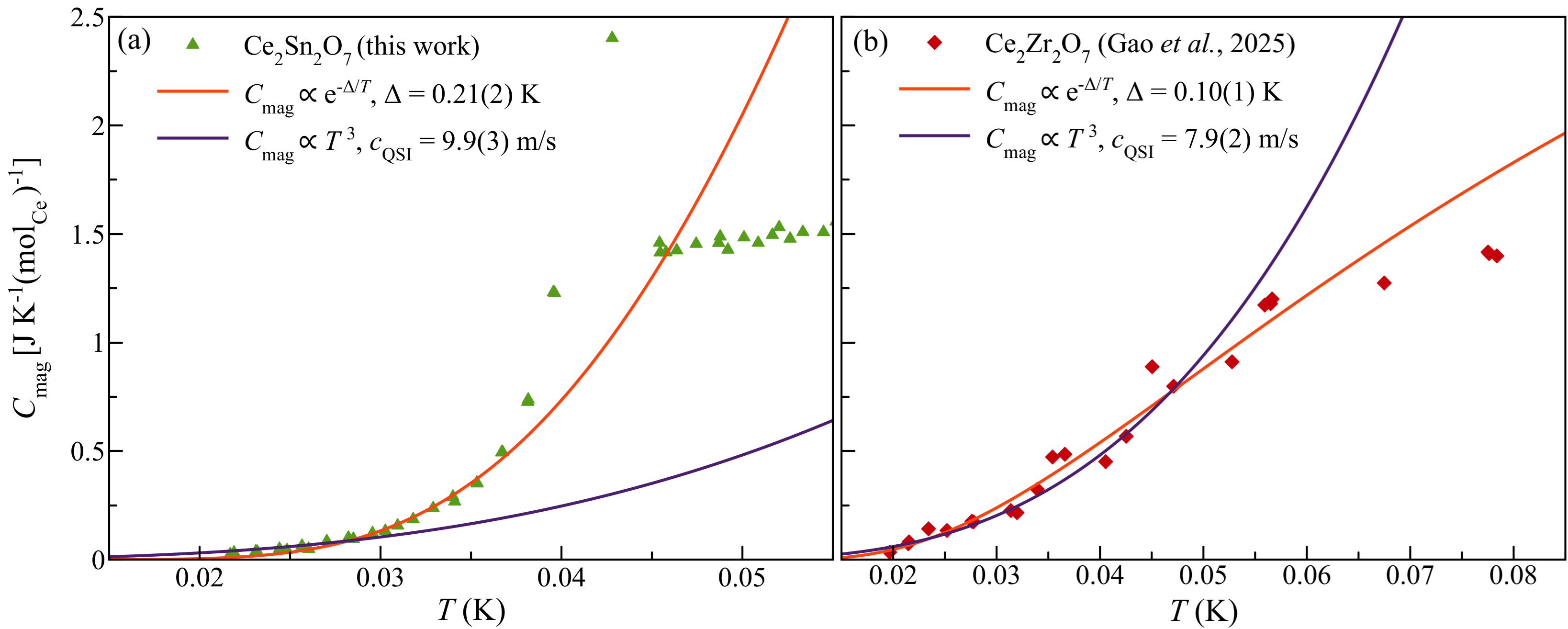}
\par
\caption{Exponential (red) and cubic (purple) extrapolations of the zero-field heat capacity to $C_{\mathrm{mag}} = 0$ at $T = 0$~K for (a) the heat capacity measured from Ce$_2$Sn$_2$O$_7$ in this work and (b) that measured from Ce$_2$Zr$_2$O$_7$ in Ref.~\cite{Gao2025}. Specifically, for Ce$_2$Sn$_2$O$_7$ we show the exponential fit obtained for the fitting range between base temperature and $T = 0.0375$~K, and the cubic fit obtained for the fitting range between base temperature and $T = 0.03$~K. For Ce$_2$Zr$_2$O$_7$ we show both fits for the fitting range between base temperature and $T = 0.055$~K. Fitting to higher temperature in each case results in an increasingly poorer description of the measure data over the fitting range.} 
\label{FigureS2}
\end{figure*}

\subsection{Low-Temperature Fits to the Heat Capacity}

In Fig.~\ref{FigureS2}, we perform fits to the low-temperature heat capacity, below the sharp peak at $T_N \sim 0.04$~K, using two possible simple extrapolation schemes: one (cubic) corresponding to gapless excitations and one (exponential) corresponding to gapped excitations. In Fig.~\ref{FigureS2}(a), we show fits of the lowest-temperature region of the zero-field heat capacity measured from Ce$_2$Sn$_2$O$_7$, below the sharp peak at $T_N \sim 0.04$~K, using these exponential ($C_{mag} \propto e^{-\Delta/T}$) and cubic ($C_{mag} \propto T^3$) forms. Figure~\ref{FigureS2}(a) shows that the exponential form fits the measured data better and over a wider temperature regime than the cubic form does. This is consistent with the expectation of a gapped excitation spectrum for the all-in all-out ordered ground state predicted for Ce$_2$Sn$_2$O$_7$ in zero field in Ref.~\cite{Yahne2024}. In Fig.~\ref{FigureS2}(a), we show the exponential form fit between base temperature and $T = 0.0375$~K. Fitting to higher temperature results in an increasingly poorer description of the measure data over the fitting range. This best fit exponential ($C_{mag} \propto e^{-\Delta/T}$) curve has a corresponding gap energy of $\Delta = 0.21 \pm 0.02$~K. In Fig.~\ref{FigureS2}(b), we show similar cubic and exponential fits of lowest-temperature region of the zero-field heat capacity measured from Ce$_2$Zr$_2$O$_7$ in Ref.~\cite{Gao2019}. As shown in Fig.~\ref{FigureS2}(b), the exponential and cubic forms fit the low-temperature heat capacity of Ce$_2$Zr$_2$O$_7$ from Ref.~\cite{Gao2019} equally well within the scatter of the data.

A cubic heat capacity below $T_N$ would be appropriate for emergent photon excitations of a QSI ground state~\cite{Benton2012, Li2017, Kato2015, Huang2020}. The corresponding emergent photon speed of light $c_{\mathrm{QSI}}$ can be estimated from the cubic ($C_{\mathrm{mag}} = AT^3$) fits to the measured heat capacity via~\cite{Benton2012, Gao2025}:
\begin{equation} \label{eq:4}
    A = \frac{R\pi^2}{60}\left(\frac{k_{\mathrm{B}}a}{\hbar c_{\mathrm{QSI}}}\right)^3
\end{equation}
where $R$ is the molar gas constant and $a$ is the cubic lattice parameter, estimated to be $a = 10.645~\angstrom$ for hydrothermally grown Ce$_2$Sn$_2$O$_7$ in Ref.~\cite{Yahne2024} and $a = 10.7~\angstrom$ for floating-zone grown Ce$_2$Zr$_2$O$_7$ in Ref.~\cite{Gao2025}. For the cubic fits to the low-temperature heat capacities of Ce$_2$Sn$_2$O$_7$ and Ce$_2$Zr$_2$O$_7$ in Fig.~\ref{FigureS2}, this gives $c_{\mathrm{QSI}} = 9.9 \pm 0.3$~m/s for Ce$_2$Sn$_2$O$_7$ and $c_{\mathrm{QSI}} = 7.9 \pm 0.2$~m/s for Ce$_2$Zr$_2$O$_7$. Interestingly, very similar speeds of light are obtained from the cubic fits to the heat capacities of Ce$_2$Sn$_2$O$_7$ and Ce$_2$Zr$_2$O$_7$ despite the fact that the magnetic ground state of Ce$_2$Sn$_2$O$_7$ is clearly ordered and an exponential form describes the measured data from Ce$_2$Sn$_2$O$_7$  better than the cubic form. In the case of Ce$_2$Zr$_2$O$_7$, the exponential form and cubic form fit the measured data equally well within the scatter of the data, as is also noted in Ref.~\cite{Gao2025}. 

\subsection{Schottky Form for the Heat Capacity of the $\alpha$ Chains}

In Fig.~4(a) of the main text we compare the heat capacity measured from single crystal Ce$_2$Sn$_2$O$_7$ in the present work with a Schottky form of the heat capacity that approximates the contribution from the $\alpha$ chains in Ce$_2$Sn$_2$O$_7$, for a magnetic field nominally along the $[1,1,0]$ direction and for field strengths between $h = 0.5$~T and 9~T. Specifically, this form for the $\alpha$-chain contribution is given by 
\begin{equation}
\label{eq:5}
C_{\alpha} = \frac{R}{2}\Big(\frac{\Delta}{2k_{\mathrm{B}}T}\Big)^2 \sech^2\Big(\frac{\Delta}{2k_{\mathrm{B}}T}\Big)
\end{equation}

\noindent where $R$ is the molar gas constant, $\Delta = g_z \mu_{\mathrm{B}} h_z$, and $h_z = h\cos(35.26^\circ)$ for a $[1,1,0]$ field direction~\cite{Hiroi2003}. This approximation takes into account only the Zeeman energy and neglects pseudospin-pseudospin interactions. Accordingly, this Schottky form is expected to describe the measured data best at high fields and this increased accuracy of the description with increasing field strength is indeed shown in Fig.~4(a).

\begin{figure*}[t]
\linespread{1}
\par
\includegraphics[width=7.2in]{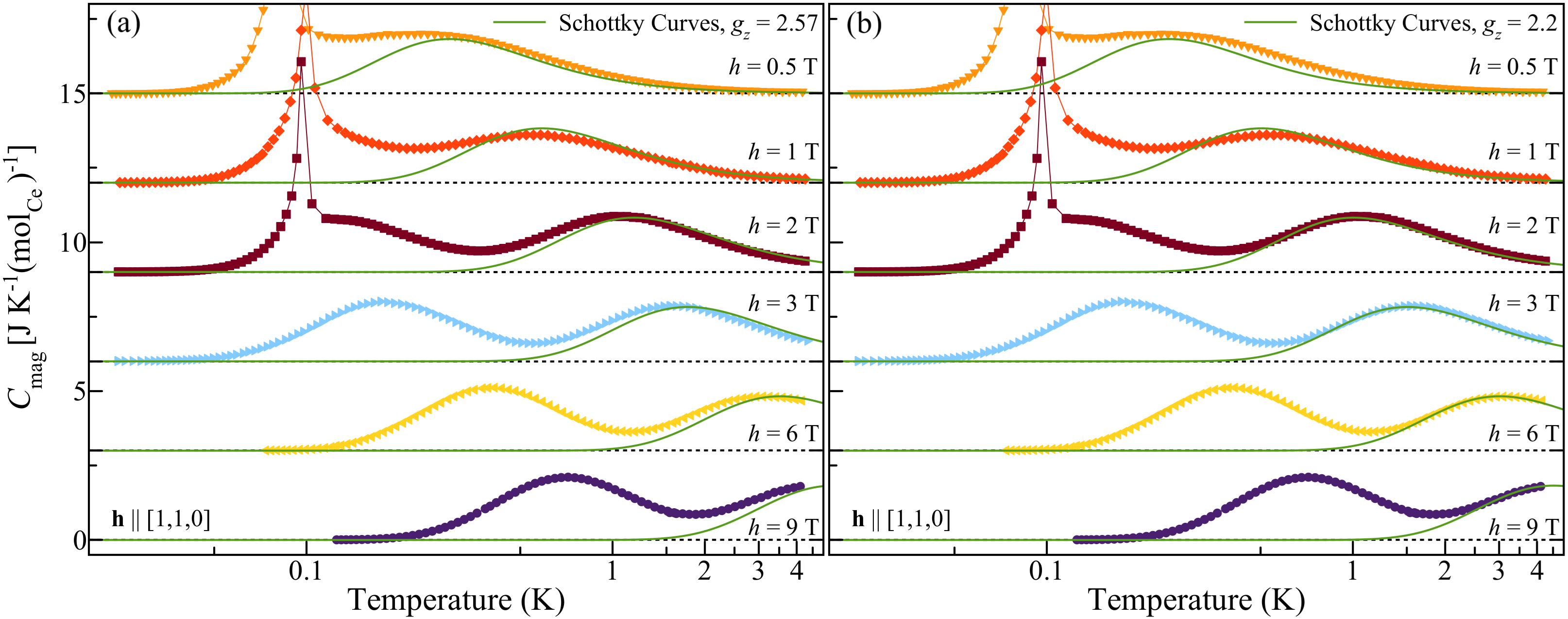}
\par
\caption{(a,b) The temperature-dependence of the heat capacity of Ce$_2$Sn$_2$O$_7$ at various field strengths (as labeled) for a magnetic field nominally along the $[1,1,0]$ direction, shown with a logarithmic temperature axis and linear heat capacity axis. We compare these datasets to the corresponding curves predicted according to the Schottky form in Eq.~\ref{eq:5} for (a)~$g_z = 2.57$ and (b)~$g_z = 2.2$. Panel~(b) is also shown in Fig.~4(a) of the main text. The data and calculations are each offset, for visibility, by 3~J~K$^{-1}$mol$_{\mathrm{Ce}}^{-1}$ between each pair of successive field strengths.} 
\label{FigureS3}
\end{figure*}

In Fig.~\ref{FigureS3}, we compare how well the Schottky form $C_{\alpha}$ describes the measure data for $C_{\alpha}$ calculated with $g_z = 2.57$ [Fig.~\ref{FigureS3}(a)] and for $C_{\alpha}$ calculated with $g_z = 2.2$ [Fig.~\ref{FigureS3}(b), also shown in Fig.~4(a)]. The value of $g_z = 2.57$ corresponds to a pure $\ket{m_J = \pm 3/2}$ CEF ground state while the estimated value of $g_z = 2.2$ (Ref.~\cite{Sibille2015, Yahne2024}) implies mixing of the $\ket{m_J = + 3/2}$ and $\ket{m_J = - 3/2}$ states in each state of the CEF ground state doublet. As shown in Fig.~\ref{FigureS3}, the Schottky form for the $\alpha$ chains provides a much better description of the data for the previously-estimated value of $g_z = 2.2$ compared to the pure $\ket{m_J = \pm 3/2}$ value of $g_z = 2.57$, particularly at higher fields where this form is expected to be most-accurate. 

\newpage
\subsection{Numerical Linked Cluster Calculations of \texorpdfstring{$C_{\mathrm{mag}}$}~}

By fitting the zero-field magnetic heat capacity, $C_{\mathrm{mag}}$, of Ce$_2$Sn$_2$O$_7$ with high-temperature NLC calculations~\cite{Rigol2006,Tang2013,Schafer2020,schaefer_magnetic_2022}, we determine the optimal nearest-neighbor coupling parameters of the pseudospin Hamiltonian. A similar fitting procedure for Cerium-based pyrochlores has been employed in Refs.~\cite{Smith2022,Yahne2024,Smith2025b}.

We parameterize the symmetry-allowed zero-field Hamiltonian at the nearest-neighbor level by introducing a permutation $\{a,b,c\}$ of $\{\tilde{x},\tilde{y},\tilde{z}\}$ such that the corresponding coupling constants satisfy $J_a \geq |J_b| \geq |J_c|$:
\begin{equation} \label{eq:6}
\begin{split}
    \mathcal{H}_\mathrm{ABC} & = \sum_{<ij>}[J_{a}{S_i}^{a}{S_j}^{a} + J_{b}{S_i}^{b}{S_j}^{b} + J_{c}{S_i}^{c}{S_j}^{c}] \\
    & = \sum_{\langle ij \rangle}[J_{a}{S_i}^{a}{S_j}^{a} - J_{\pm}({S_i}^{+}{S_j}^{-} + {S_i}^{-}{S_j}^{+}) \\
    & + J_{\pm\pm}({S_i}^{+}{S_j}^{+} + {S_i}^{-}{S_j}^{-})] \;,
\end{split}
\end{equation}
where $J_\pm = -\frac{1}{4}(J_b + J_c)$ and $J_{\pm\pm} = \frac{1}{4}(J_b - J_c)$.

The best-fit parameters determine whether the corresponding zero-field ground state of the Hamiltonian is ordered or disordered. However, the zero-field heat capacity is not susceptible to either the permutation of the pseudospin axes or the mixing angle $\theta$ defined in Eqs.~2 and 3 of the main text. Consequently, it does not determine whether the ground state is of dipolar or octupolar character.

We consider 2278 uniformly distributed sets of coupling parameters and compute the heat capacity using a sixth-order NLC expansion, including all translationally inequivalent subclusters containing up to six corner-sharing tetrahedra. We use the Euler extrapolation technique to improve convergence~\cite{Tang2013,Schafer2020,schaefer_magnetic_2022}. For each parameter set, we evaluate the residual with respect to the experimental data.
We first fit the high-temperature data in the range $1.5$~K~$\leq T \leq 6$ K to determine the overall energy scale. Keeping this scale fixed, we then evaluate the residual using all experimental data points in the temperature range $0.15$~K~$\leq T_{\mathrm{exp}} \leq 1.5$ K:
\begin{equation}\label{eq:7}
\delta^2 = \sum_{T_{\mathrm{exp}}}
\left[
C_{\mathrm{mag}}^{\mathrm{NLC},6}(T_{\mathrm{exp}})
-
C_{\mathrm{mag}}^{\mathrm{exp}}(T_{\mathrm{exp}})
\right]^2 ,
\end{equation}
\noindent where $C_{\mathrm{mag}}^\mathrm{exp}(T_\mathrm{exp})$ is the measured heat capacity at temperature $T_\mathrm{exp}$ and $C_\mathrm{mag}^{\mathrm{NLC},6}(T_\mathrm{exp})$ is the heat capacity calculated via the sixth-order NLC method at temperature $T_\mathrm{exp}$.


\begin{figure*}[t]
\linespread{1}
\par
\includegraphics[width=7.2in]{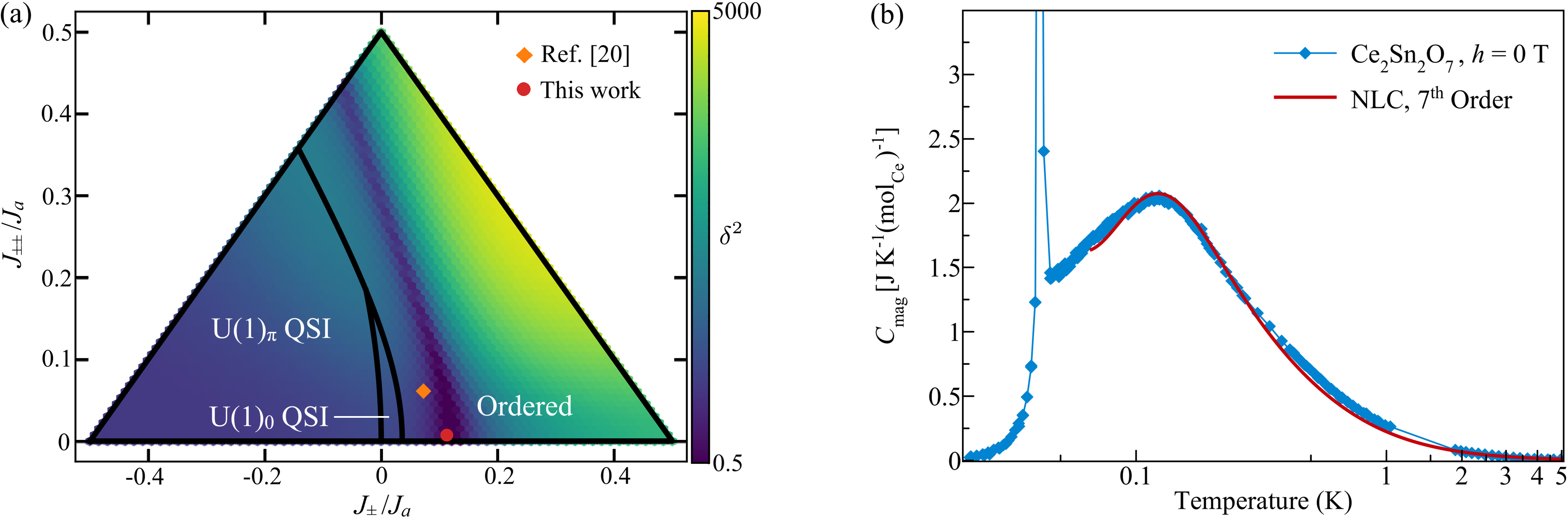}
\par
\caption{(a) The goodness-of-fit parameter $\delta^2$ for our sixth-order NLC fits to the measured $C_\mathrm{mag}$ of Ce$_2$Hf$_2$O$_7$, shown on a logarithmic scale. We also show the phase boundaries and corresponding phases in the ground state phase diagram predicted at the nearest-neighbor level for dipole-octupole pyrochlores~\cite{Benton2020}. The best-fit parameters from this fitting procedure are marked as a red circle and the best-fit parameters from the similar fitting procedure in Ref.~\cite{Yahne2024} are marked as an orange diamond. (b)~The $C_\mathrm{mag}$ measured from single crystal Ce$_2$Hf$_2$O$_7$ in this work as well as the $C_\mathrm{mag}$ calculated via seventh-order NLC using the parameters at the U(1) symmetric point $(J_a,J_b,J_c) = (0.038, -0.009, -0.009)$~meV near the best-fit parameters obtained from our fitting procedure, $(J_a,J_b,J_c) = (0.038, -0.008, -0.009)$~meV.} 
\label{FigureS4}
\end{figure*}


The residual is shown in Fig.~S4(a) as a function of $J_\pm/J_a$ and $J_{\pm\pm}/J_a$ for all symmetry-inequivalent parameter sets. Parameter combinations within the white regions correspond to different permutations of the coupling strengths that cannot be distinguished by the zero-field heat capacity. Blue (yellow) indicates better (worse) agreement with the experimental heat capacity.
As observed for other cerium pyrochlore compounds~\cite{Smith2022,Smith2025b,Yahne2024}, we find a well-defined isolated valley that provides a good description of the experimental data. The best-fit parameters, $(J_a,J_b,J_c) = (0.038,-0.008,-0.009)$ meV, lie close to the U(1)-symmetric point with $J_{\pm\pm}/J_a \approx 0$.
To assess the convergence of the seventh-order NLC expansion, we perform a seventh-order calculation using the U(1)-symmetrized parameter set obtained by replacing $J_b$ and $J_c$ with their average, $(J_b+J_c)/2$. The resulting heat capacity is shown in Fig.~S4(b). The sixth- and seventh-order NLC calculations are in agreement down to $T \sim 0.07$ K, demonstrating convergence over this temperature range. The calculation accurately reproduces the Schottky anomaly and loses convergence only upon approaching the temperature of the experimentally observed phase transition. 

The resulting parameters are similar to those obtained in Ref.~\cite{Yahne2024}, $(J_a,J_b,J_c) = (0.045, -0.001, -0.012)$~meV, with both sets of parameters lying in a region of parameter space where a spin ice phase is classically stable in zero field but is ultimately unstable to an ordered zero-field ground state when quantum effects are considered.
Notably, the heat capacity measurements in Ref.~\cite{Yahne2024} had a large 13\% systematic uncertainty associated with the small sample mass, leading to an additional parameter scaling the measured heat capacity used for the fitting process, while the systematic uncertainty associated with the measurements of the present work is $\sim 0.1$\% and is considered to be negligible along with any small unsystematic uncertainty. 

\newpage
\section{Nearest-Neighbor XYZ Hamiltonian for \texorpdfstring{$\mathrm{Ce}_2\mathrm{Sn}_2\mathrm{O}_7$}~}

The pseudospin transformations between the $\{x,y,z\}$ and $\{\tilde{x},\tilde{y},\tilde{z}\}$ local coordinate frames used for Eqs.~1 and 3 of the main text are:
\begin{equation} 
    {\hat{S}_i}^{\tilde{x}} = {\hat{S}_i}^{x}\cos{\theta} + {\hat{S}_i}^{z}\sin{\theta},
\end{equation}
\begin{equation} 
    {\hat{S}_i}^{\tilde{z}} =  -{\hat{S}_i}^{x}\sin{\theta} + {\hat{S}_i}^{z}\cos{\theta},
\end{equation}

\noindent or in the other direction:

\begin{equation} 
    {\hat{S}_i}^{x} = {\hat{S}_i}^{\tilde{x}}\cos{\theta} - {\hat{S}_i}^{\tilde{z}}\sin{\theta},
\end{equation}
\begin{equation} 
    {\hat{S}_i}^{z} = {\hat{S}_i}^{\tilde{x}}\sin{\theta} +  {\hat{S}_i}^{\tilde{z}}\cos{\theta}
\end{equation}

\noindent with ${\hat{S}_i}^{y} = {\hat{S}_i}^{\tilde{y}}$. \\


\begin{table}[!h]
\begin{center}
\begin{tabular}{|c|c|c|c|c|}
\hline
$\gamma$ & 1 & 2 & 3 & 4 \\\hline
\renewcommand{\arraystretch}{2}
\begin{tabular}[c]{@{}c@{}} $\mathbf{x}_\gamma$  \end{tabular}          
& $\frac{1}{\sqrt{6}}[-2,1,1]$ & $\frac{1}{\sqrt{6}}[-2,-1,-1]$ & $\frac{1}{\sqrt{6}}[2,1,-1]$ & $\frac{1}{\sqrt{6}}[2,-1,1]$ \\ \hline
\renewcommand{\arraystretch}{2}
\begin{tabular}[c]{@{}c@{}} $\mathbf{y}_\gamma$ \end{tabular}          
& $\frac{1}{\sqrt{2}}[0,-1,1]$ & $\frac{1}{\sqrt{2}}[0,1,-1]$ & $\frac{1}{\sqrt{2}}[0,-1,-1]$ & $\frac{1}{\sqrt{2}}[0,1,1]$ \\ \hline
\renewcommand{\arraystretch}{2}
\begin{tabular}[c]{@{}c@{}} $\mathbf{z}_\gamma$  \end{tabular}          
& $\frac{1}{\sqrt{3}}[1,1,1]$ & $\frac{1}{\sqrt{3}}[1,-1,-1]$ & $\frac{1}{\sqrt{3}}[-1,1,-1]$ & $\frac{1}{\sqrt{3}}[-1,-1,1]$ \\ \hline
\end{tabular}
\caption{The local $x$, $y$, and $z$ directions with respect to the cubic crystallographic axes for the four different sublattices ($\gamma = 1,2,3,4$) composing the tetrahedral network of rare-earth ions in the $R_2B_2$O$_7$ pyrochlores~\cite{Ross2011, SmithThesis}. In this sublattice description, the location of each $R^{3+}$ ion with respect to the center of its tetrahedron is given by $\mathbf{b}_\gamma = -\frac{\sqrt{3}}{8} \mathbf{z}_\gamma$, and these tetrahedra form a face-centered cubic lattice with tetrahedra centered on the primitive lattice locations, $R = n_1\mathbf{a}_1 + n_2\mathbf{a}_2 + n_3\mathbf{a}_3$ for $n_1, n_2, n_3 \in \mathbb{Z}$, where $\mathbf{a}_1 = \frac{a}{2}[1/2,1/2,0]$, $\mathbf{a}_2 = \frac{a}{2}[1/2,0,1/2]$, and $\mathbf{a}_3 = \frac{a}{2}[0,1/2,1/2]$. For a magnetic field nominally along $[1,1,0]$, the $\gamma = 1,4$ ($\gamma = 2,3$) sites are within the $\alpha$ chains ($\beta$ chains) of the pyrochlore lattice.}
\label{Tab:1}
\end{center}
\end{table}

In terms of the local axes for the four rare-earth sublattices (see Table I), these transformations correspond to $\mathbf{\tilde{x}}_\gamma = (\cos\theta)\mathbf{x}_\gamma + (\sin\theta)\mathbf{z}_\gamma$, $\mathbf{\tilde{y}}_\gamma = \mathbf{y}_\gamma$, and $\mathbf{\tilde{z}}_\gamma = -(\sin\theta)\mathbf{x}_\gamma + (\cos\theta)\mathbf{z}_\gamma$, for $\gamma = 1,2,3,4$. Using the estimated value of $\theta = 0.19\pi$ for Ce$_2$Sn$_2$O$_7$ from Ref.~\cite{Yahne2024}, together with $\mathbf{x}_\gamma$, $\mathbf{y}_\gamma$, and $\mathbf{z}_\gamma$ in Table~I, gives the local $\mathbf{\tilde{x}}_\gamma$, $\mathbf{\tilde{y}}_\gamma$, and $\mathbf{\tilde{z}}_\gamma$ axes listed in Table~II. 

\begin{table*}[]
\begin{center}
\begin{tabular}{|c|c|c|c|c|}
\hline
$\gamma$ & 1 & 2 & 3 & 4 \\\hline
\renewcommand{\arraystretch}{2}
\begin{tabular}[c]{@{}c@{}} $\mathbf{\tilde{x}}_\gamma$  \end{tabular}          
& $[-0.3508,0.6621,0.6621]$ & $[-0.3508, -0.6621,-0.6621]$ & $[0.3508,0.6621,-0.6621]$ & $[0.3508,-0.6621,0.6621]$ \\ \hline
\renewcommand{\arraystretch}{2}
\begin{tabular}[c]{@{}c@{}} $\mathbf{\tilde{y}}_\gamma$ \end{tabular}          
& $\frac{1}{\sqrt{2}}[0,-1,1]$ & $\frac{1}{\sqrt{2}}[0,1,-1]$ & $\frac{1}{\sqrt{2}}[0,-1,-1]$ & $\frac{1}{\sqrt{2}}[0,1,1]$ \\ \hline
\renewcommand{\arraystretch}{2}
\begin{tabular}[c]{@{}c@{}} $\mathbf{\tilde{z}}_\gamma$  \end{tabular}          
& $[0.9365,0.2480,0.2480]$ & $[0.9365,-0.2480,-0.2480]$ & $[-0.9365,0.2480,-0.2480]$ & $[-0.9365,-0.2480,0.2480]$ \\ \hline
\end{tabular}
\caption{The local $\tilde{x}$, $\tilde{y}$, and $\tilde{z}$ directions with respect to the cubic crystallographic axes for the four different Ce$^{3+}$ sublattices ($\gamma = 1,2,3,4$) in Ce$_2$Sn$_2$O$_7$ based on the estimated value of $\theta = 0.19\pi$ for Ce$_2$Sn$_2$O$_7$ in Ref.~\cite{Yahne2024}.}
\label{Tab:2}
\end{center}
\end{table*}

According to the XYZ Hamiltonian parameters for Ce$_2$Sn$_2$O$_7$ in Ref.~\cite{Yahne2024}, $(J_{\tilde{x}}, J_{\tilde{y}}, J_{\tilde{z}}) = (0.045, -0.001, -0.012)$~meV and $\theta = 0.19\pi$, the ground state of the XYZ Hamiltonian for Ce$_2$Sn$_2$O$_7$ is an ${S}^{\tilde{z}}$-flavored all-in all-out ordered phase in which the pseudospins are either all along their respective local $+\tilde{z}$ directions or all along their respective local $-\tilde{z}$ directions. This phase is stabilized by quantum fluctuations and in the classical phase diagram the ground state is instead a ${S}^{\tilde{x}}$-flavored spin ice phase in which two pseudospins on each tetrahedron point along their local $+\tilde{x}$ directions and two point along their local $-\tilde{x}$ directions. However, as mentioned in the main text, recent diffuse neutron scattering results on hydrothermally grown Ce$_2$Sn$_2$O$_7$ in Ref.~\cite{Yuan2026} suggest that long-ranged dipole-dipole interactions must be accounted for to achieve an adequate description of the low-temperature magnetic behavior in Ce$_2$Sn$_2$O$_7$.

Here we also discuss some of the phases that are relevant to the $\beta$ chains in Ce$_2$Sn$_2$O$_7$ in a magnetic field nominally along $[1,1,0]$, according to the XYZ Hamiltonian parameters for Ce$_2$Sn$_2$O$_7$ in Ref.~\cite{Yahne2024}. For the $[1,1,0]$ field direction, it is useful to decompose the pyrochlore lattice into $\alpha$ and $\beta$ chains extending along the $[1,1,0]$ and $[1,\bar{1},0]$ directions, respectively. This is particularly convenient as the ions in the $\beta$-chains have their dipole moments (their local $\pm z$ directions) perpendicular to the $[1,1,0]$ field direction, leaving them essentially decoupled from the field. On the other hand, dipole moments in the $\alpha$-chains each have a component along the $[1,1,0]$ field direction, leading to non-collinearly field-polarized $\alpha$-chains with pseudospins along their local $\pm z$ directions as to give a net magnetic moment that is along the field direction. Further simplifying the picture is that the $\beta$ chains are decoupled from the $\alpha$ chains when the $\alpha$ chains are polarized~\cite{Yoshida2004, Placke2020}. This is a result of magnetic frustration associated with the fact that, when the $\alpha$ chains are polarized, each $\beta$-chain ion is nearest-neighbors with two $\alpha$-chain ions that have their pseudospin along $+z$ and two $\alpha$-chain ions that have their pseudospin along $-z$, leading to a cancellation of the interactions. The fact that the parameters from Ref.~\cite{Yahne2024} satisfy $J_{\tilde{x}} \gg |J_{\tilde{y}}|$,$|J_{\tilde{z}}| > 0$ suggests that antiferromagnetic intrachain correlations between ${S}^{\tilde{x}}$ components of pseudospin would be the dominant correlation in the $\beta$ chains of Ce$_2$Sn$_2$O$_7$ when the $[1,1,0]$-field has polarized the $\alpha$ chains. These correlations are ferromagnetic with respect to a \textit{global} coordinate system. Accordingly, it is expected that even a small misalignment of the magnetic field away from the $[1,1,0]$ direction leads to a significant coupling between the magnetic field and the $\beta$ chains, as has been found for Ce$_2$Hf$_2$O$_7$ in Ref.~\cite{Bhardwaj2025}. Furthermore, in light of the diffuse neutron scattering work in Ref.~\cite{Yuan2026} highlighting the significance of long-ranged dipole-dipole interactions in Ce$_2$Sn$_2$O$_7$, it is likely that dipole-dipole interactions play a large role in any interchain coupling of the $\beta$ chains in moderate $[1,1,0]$ magnetic fields at low temperature.

\end{document}